\documentclass[11pt]{article}

\input epsf.sty
\usepackage{color} %used for font color
\usepackage[utf8]{inputenc} %useful to type directly accentuated characters
\usepackage{feynmp}

\usepackage{color}
\definecolor{red}{cmyk}{0,1,0.50002,0}
\definecolor{blue}{rgb}{0,0.49995,1}
\definecolor{green}{cmyk}{1,0,0.49998,0}
 
\usepackage{graphics,amsmath,amssymb}

\def\lromn#1{\uppercase\expandafter{\romannumeral#1}}

\begin{document}

\begin{center}
\begin{large}
\textbf{
Gravitational and dark wave emission
at binary merger
}
\end{large}
\end{center}

\begin{center}
\begin{large}

Kunio Kaneta\footnote{kkaneta@lab.twcu.ac.jp},
Kin-ya Oda\footnote{odakin@lab.twcu.ac.jp}, 
and Motohiko Yoshimura$^{\dagger}$\footnote{yoshim@okayama-u.ac.jp} 

\vspace{0.5cm}
Department of Mathematics, 
Tokyo Woman's Christian University, \\
Tokyo 167-8585, Japan

\vspace{0.5cm}
$^{\dagger}$
Research Institute for Interdisciplinary Science,
Okayama University, \\
Tsushima-naka 3-1-1 Kita-ku Okayama
700-8530, Japan

\end{large}
\end{center}

\vspace{3cm}

%\vspace{0.5cm}
\begin{center}
\begin{Large}
{\bf ABSTRACT}
\end{Large}
\end{center}

The recently proposed formalism of extended Jordan-Brans-Dicke  gravity
makes it possible to calculate energy loss rate
due to both gravitational wave  and scalar field 
(giving the origin of dark energy) wave  emission
at merger of a black hole  and a neutron star;
a binary system of no scalar hair and a star with the scalar charge.
The scalar field emission changes orbit parameters
of the binary system, thereby changes detectable gravitational wave emission.
When neutron stars carry significantly large scalar  charge, significant
dark wave (namely, scalar field wave) emission 
occurs at the same time of gravitational wave emission.
It is found that solutions of coupled differential
equations predict non-vanishing remnant dark charge after the gravitational collapse.
This gives two interesting possibilities:
(1) the no-hair conjecture of black hole is violated, or
(2)
a bosonic cloud  is formed outside the event horizon of black hole.
The bosonic cloud proposed in the literature is a gigantic atom made of gravitationally bound
 dark energy quanta surrounding a black hole. 
One can either constrain, or even determine, parameters of extended
Jordan-Brans-Dicke gravity from accumulated gravitational wave observations
of merging black hole and neutron star.
\footnote{
Key words

Gravitational wave emission,
Dark energy,
No hair theorem,
Jordan-Brans-Dicke gravity,
Black hole merger with neutron star,
bosonic cloud
}

\setcounter{footnote}{0}

%\begin{enumerate} \item 
\vspace{2cm}
{\bf  Introduction}

General relativity (GR) is the leading theory of gravity,
passing all tests in the weak gravity regime \cite{gr tests}, 
 and test at the initial stage
of gravitational wave (GW) emission in the strong gravity regime \cite{gw gr-test,
gw gr-test: bh-ns}.
There is however a good reason to consider scalar-tensor gravity
as a powerful alternative,
due to recent developments in cosmology; the dark energy problem and inflation.

A class of scalar-tensor gravity, an extended Jordan-Brans-Dicke (eJBD)
theory, has recently been identified \cite{ejbd constraints},
which pass stringent observational constraints  on
time variation of fundamental parameters.
These parameters are time-invariants in GR.
So far no deviation from GR has been found, giving constraints on eJBD theory.
Concordance of calculated energy budget evolution 
with cosmological observations found in \cite{ejbd constraints}
 suggests that the time varying mass of eJBD field is small,
of order the Hubble rate $\sim 10^{-33}$ eV 
 at the present epoch, thus making the original massless JBD theory \cite{jbd},
(however modified by its coupling and finite potential), a relevant theory for analysis
at recent epochs.

In the present work we discuss GW emission at binary merger in JBD gravity.
 The energy loss  due to the scalar field emission accompanying GW emission may
become comparable to, or even dominate over, GW;
it may thus drastically change result of pure GW amplitude signals in GR.
BH-BH merger proceeds via pure GW emission adopting the no-hair conjecture.
It is thus promising to examine GW emission at
 BH- neutron star (NS) merger, since NS is allowed to have the scalar charge
called the dark charge, too.
Discussion of GW emission rate at 
NS-NS merger is difficult due to increased calculational uncertainty.
Our simple analysis of BH-NS merger presented in this work may
enable one to place useful constraint on model parameters in eJBD gravity
if GW detection data from BH-NS merger are accumulated.
One needs  to introduce two-component O(2) symmetric  eJBD fields
in order to define a conserved dark charge unambiguously.

We use the natural unit of
$c=1 = \hbar$ unit unless otherwise stated.
A typical energy loss rate of gravitational wave emission
is 1 solar mass/ msec $= 1.8 \times 10^{57}$ erg/sec.

\vspace{0.5cm}
{\bf  Static solution with scalar hair}

We start our discussion from the no-hair conjecture of isolated compact objects.
Spherically symmetric vacuum solutions  of JBD gravity are well known;
for example \cite{jbd static sol}.
Their metric and JBD field are given by
\begin{eqnarray}
&&
ds^2 = f(r) dt^2 - \frac{dr^2}{f(r)} - g(r) (d\theta^2 + \sin^2 \theta d\phi^2)
\,,
\label {jbd metric}
\\ &&
f(r) = (1- \frac{l}{r})^{\alpha}
\,, \hspace{0.5cm}
g(r) = r^2 (1- \frac{l}{r})^{1-\alpha}
\,,
\label {jbd metric 2}
\\ &&
\chi(r) = \chi(\infty) + \frac{q M}{2} \ln (1- \frac{l}{r})
\,, \hspace{0.5cm}
\alpha^2 + q^2 =1
\,.
\label {jbd metric 3}
\end{eqnarray}
$G$ is the gravitational constant.
The mass $M$ and the JBD charge $D$, called the dark charge hereafter,
 at the center of massive body are defined from asymptotic behaviors
at $r \rightarrow \infty$:
\begin{eqnarray}
&&
g_{tt} = 1 - \frac{2GM}{r} + \cdots
\,, \hspace{0.5cm}
\chi = \chi(\infty) - \frac{GD M}{r} + \cdots
\,.
\label {asymptotic behavior}
\end{eqnarray}
Parameters of the solution in (\ref{jbd metric}) $\sim$ (\ref{jbd metric 3})
 are given in terms of $M$ and $D$;
\begin{eqnarray}
&&
l = 2 G \sqrt{M^2+ D^2}
\,, \hspace{0.3cm}
q = \frac{D}{\sqrt{M^2+ D^2}}
\,, \hspace{0.3cm}
\alpha = \frac{M}{\sqrt{M^2+ D^2}}
\,.
\end{eqnarray}

Divergence  in the metric (\ref{jbd metric})  at $r=l$ gives a curvature singularity with
the scalar quantity $R_{\mu \nu \rho \sigma}R^{\mu \nu \rho \sigma} = \infty$:
it is a naked singularity not covered by event horizon.
The no-hair conjecture, often called no-hair theorem in the literature,
 states that physically allowed solution at gravitational collapse
is black hole without the curvature singularity.
\footnote{
Unstable linearized perturbation modes
around general static solution having the curvature
singularity were found \cite{jbd instability}.
Since  dynamical  gravitational collapse proceeds without
developing unstable modes, this suggests that 
formation of black hole ends  without the
curvature singularity; namely, the Schwarzschild metric is formed,
as discussed in  \cite{jbd instability}.
This argument may be taken as a support for no-hair conjecture.
}
The Schwarzshild solution defined by the condition $\alpha=1\,, q =0$ (hence $D=0$),
 is thus the unique static solution giving a candidate black hole.
In an isolated NS or a white dwarf a finite dark charge  may exist with
$D \neq 0$. In this case neither  the event horizon
nor  the curvature singularity is present, since the Schwarzschild radius
is within the neutron star surface where the above vacuum solution
becomes irrelevant, the presence of matter precluding the event horizon
and the curvature singularity.

\vspace{0.5cm}
%\item
 {\bf Scalar hair in eJBD gravity}

We introduce two-component real eJBD fields:
$\chi = \chi_1 + i \chi_2 = \chi_r e^{i \theta}$ with real
$\chi_i (i = 1,2,r)$ and $\theta$ fields.
In models of eJBD gravity that we consider in the present work
the   spatially homogeneous background of  $\chi_r$ evolves with cosmological
evolution, and  near the present cosmic epoch it is given by
$\chi_r = \chi_0(t) \sim A t^{1/\delta} \,, \delta > 1 $,
with $A\,, \delta $ two constants. 
Time variation arises from a non-trivial potential in the  reformulated eJBD theory
\cite{ejbd constraints}.
The dark eJBD field $\chi$ couples to  matter fields of particle physics
(fermions, gauge bosons, and Higgs boson) via interaction lagrangian 
${\cal L}^{\rm E}_{st}$ in the Einstein metric frame,
\begin{eqnarray}
&&
{\cal L}^{\rm E}_{st} =  F^{p_g}{\cal L}_g + F^{p_{\rm f} } {\cal L}_f -
F^{p_H} V_{\rm J}(H) +   F^{p_{\rm dH}} (D_{\mu} H)^{\dagger} (D^{\mu} H)
+ F^{p_Y}{\cal L}_y
\nonumber \\ &&
\hspace*{1cm}
+ i \frac{1}{2} p_f\, \sum_{i = 1,2} \partial_{\chi_i}  \ln F \, \partial_{\mu} \chi_i \,
(3 j_B^{\mu} + j_L^{\mu} )
\,,
\label {lagrangian in e-frame}
\end{eqnarray}
where ${\cal L}_g$ is SU(3)$\times$ SU(2)$\times $ U(1) gauge field
contribution, and ${\cal L}_f $ refers to kinetic lagrangian of fermions, while
${\cal L}_y$ is Yukawa coupling of fermion to Higgs boson,
other ${\cal L}_i$'s being Higgs potential and Higgs kinetic lagrangian.
$3 j_B^{\mu} + j_L^{\mu} = \sum_f \bar{f} \gamma^{\mu} f$
is the universal fermion current summed over all fundamental fermions $f$.
The conformal function is taken as $F(\chi) = 1+ c (\chi_r/M_{\rm P})^2 $ with $c>0$ 
a constant,\footnote{
Most generally, there can be five independent conformal functions and
five independent powers \cite{ejbd constraints} consistent with low energy limit
of super-string theory \cite{damour-polyakov}.
We simplified this general case by introducing a common conformal function $F$.
}
 and  $M_{\rm P} =1/\sqrt{8\pi G}
\sim 2.4 \times 10^{18}\, {\rm GeV}$ the reduced Planck mass.
The lagrangian density (\ref{lagrangian in e-frame}) is a function of eJBD fields,
$\chi_r^2$ and $\partial \chi_r^2$, while its kinetic terms contain
$(\partial \chi_r)^2$ and $(\chi_r \partial \theta)^2 $.

Spatially inhomogeneous eJBD field develops inside compact objects such
as neutron star and white dwarf where many nucleons and electrons act
as the source term of $\chi$ field.
Relevant lagrantian parts in (\ref{lagrangian in e-frame})
for the present work  are nucleon and electron mass terms
and coupling term in the second line of (\ref {lagrangian in e-frame}),
since these make up ordinary matter of a massive astrophysical body.
The main nucleon mass originates from quantum chromo-dynamic interaction
given in ${\cal L}_g$, with a small additional contribution of u- and d-quark masses
arising from ${\cal L}_y$ in the spontaneously broken phase.

The power factors $p_i$ and the
conformal function $F(\chi)$ are expected to be derivable from
super-string theory at a far more advanced stage of theoretical developments,
but at present it is completely unknown except at the string tree level
\cite{damour-polyakov}.
Null-observation on time variation of standard particle physics parameters
places bounds on the powers of conformal functions,
of order
$|p_g| /\delta < 10^{-7} \,, |p_Y|/\delta < 10^{-2} \,, \delta >1\,, |p_Y| < 1/2$,
from independent observations and laboratory tests \cite{ejbd constraints}.

The dark charge is unambiguously defined in terms of non-trivial Nambu-Goldsone
mode \cite{bifurcated ejbd}.
O(2) symmetry of eJBD $\chi$ field is spontaneously broken inside compact objects by
chameleon effect \cite{chameleon}, and the dark charge $Q_i^{\chi}$ of compact object $i$
is given by  the space-integrated time component
of the conserved current $J_{\mu}$,
\begin{eqnarray}
&&
J_{\mu} =
- i \frac{1}{2}(\chi^{\dagger} \partial_{\mu} \chi - \partial_{\mu}\chi^{\dagger}\chi)
= \chi_1 \partial_{\mu} \chi_2 - \chi_2 \partial_{\mu} \chi_1 = \chi_r^2 \partial_{\mu} 
\theta
\,, \hspace{0.5cm}
Q_i^{\chi} = \int_i d^3 x \, J_0 
\,,
\end{eqnarray} 
inside a compact object $i$.
The field equation for the angular $\theta-$part, or equivalently the conservation law,
 $\partial^{\mu} J_{\mu} = 0$, 
makes it possible to have non-vanishing, time dependent $Q_i^{\chi}$
along with non-vanishing $\vec{\nabla} \cdot \vec{J} $ inside compact objects
such that the Gauss-law holds.

On the other hand, $D$ that appears in (\ref{asymptotic behavior}) is
derived by solving $\chi$ field equation,
\begin{eqnarray}
&&
\frac{1}{\sqrt{-g}} \partial_{\mu} \partial_{\nu} (\sqrt{-g} g^{\mu\nu} \delta \chi)
=  \langle \left( \frac{\delta^2 {\cal L}^{\rm E}_{st}} {\delta \chi^2 } \right)_{\chi = \chi_0} 
\rangle_i \delta \chi 
\,,
\label {chi-field eq}
\\ &&
\left( \frac{\delta^2 {\cal L}^{\rm E}_{st}} {\delta \chi^2 } \right)_{\chi = \chi_0} 
\simeq \frac{c}{M_{\rm P}^2} \left( 2 p_g {\cal L}_{\rm gluon} - (2p_Y + p_f) 
\sum_{f=u,d,e} m_f \bar{f} f + 2 p_f \sum_{f=u,d,e} \partial \cdot \bar{f} i \gamma   f 
\right)
\,,
\label {rhs chi-field eq}
\end{eqnarray}
with $\chi = \chi_0 + \delta \chi$.
The quantity $\langle \cdots \rangle_i$ is the expectation value integrated over
a compact object $i$.
The last term in (\ref{rhs chi-field eq}) is derived by a partial integration, and
vanishes if one uses the Dirac equation $i \partial \cdot \gamma f = m_f f $, 
namely on-the-mass-shell condition.
The metric here is taken the same as in (\ref{jbd metric}).
The right hand side quantity of (\ref{chi-field eq})
 is non-vanishing only inside compact objects.
Spatial integration of this quantity gives a quantity of order,
$2c \bar{p} M_i Q^{\chi}_i $ with $\bar{p}$ an average of $p_i$ factors.
We define a quantity $\eta_0$ by $\eta_0 = 2 c \bar{p}$.
If one chooses $p_g = p_Y = 0$ in order to maintain the simplest consistency with
the null-observation of time variation, 
the right hand side of (\ref {rhs chi-field eq}) becomes 
$- c p_f/M_{\rm P}^2 $ times $\sum_{f=u,d} m_f \bar{f}f $,
which gives a fraction of quark mass contribution  $\sim 0.01$
of neutron star.
The important parameter $Q^{\chi}_{i}$ in the case of $p_g = p_Y = 0$ is of order $0.01/2$
and $\eta_0 = - c p_f$ for the neutron star.

As in the literature \cite{jbd emission}, we assume that black holes have no hair,
while neutron stars, or any other astrophysical bodies without event horizon, can have
dark charges $D $.
The purpose of the present work is how to derive
constraint on $D/M_i = O(2c \bar{p} Q_i^{\chi})$ 
from gravitational emission from BH-NS merger,
assuming that the scalar charge emission is impossible to observe by
gravitational wave detectors presently available.
If one can constraint  these power factors from GW observations,
one would have hints on low energy limit of string theories.

\vspace{0.5cm}
%\item 
{\bf Gravitational and dark wave emission from binary system}

Formulas of quadrupole gravitational wave (denoted by $GW$) emission and
dipole dark wave (denoted by $DW$) emission  rates are given in
\cite{jbd emission}.\footnote{
The formula of dark wave emission given in this paper is
based on the work \cite{dw emission rate}:}
\begin{eqnarray}
&&
L_{GW} = \frac{32}{5} \frac{ G^4 \mu^2 M^3}{ R^5} \, f_g(e)
\,, \hspace{0.3cm}
L_{DW} = \frac{G^3 \mu^2 D^2 }{R^4}  \, f_d(e) \eta_0^2
\,,
\\ &&
 f_g(e) = \frac{1+ 73 e^2/24 + 37 e^4/96 }{(1-e^2)^{7/2} }
\,, \hspace{0.3cm}
f_d (e) = \frac{ 1+ e^2/2}{(1-e^2)^{5/2} }
\,,
\end{eqnarray}
where $e$ is eccentricity, $\Omega = 2\pi/P$ with $P$ the orbital period, 
\footnote{
According to \cite{gr textbook 2}
the orbital angular momentum $J$ changes following a relation
$J^2 = GM\mu^2 (1-e^2)\, R \propto R$ for a constant eccentricity.
Remnant compact object can be rotating BH described by a Kerr solution.
}
and 
$M= M_1+M_2\,, \mu = M_1 M_2/M$ (total and reduced masses).
$\eta_0 $ is the quantity of order $2c \bar{p}$.
We assume that the eccentricity is unchanged during the entire emission.
In the case of NS-NS merger the energy loss rate is proportional to
$\eta_0^4 (D_1- D_2)^2/M_{NS}^2$.

We consider BH-NS merger as the binary system.
The effective gravitational mass $M_g$ is given by
$M_g = M_1 + M_2 - G M_1 M_2/(2R)$ in the binary system:
this formula is valid in non-relativistic Keplerian motions.
When the radius $R$ is much larger than the Schwarzshild radius $2G M_g$,
variation of gravitational mass is given by variation of $R$.
Hence one has a differential equation for the radius $R$ change:
\begin{eqnarray}
&&
\frac{dR}{dt} = - \frac{64}{5} \frac{G^3 \mu M^2 }{R^3 } f_g(e) - 
\frac{2 G^2 M_{BH}  D^2}{  M_{NS}  R^2 } f_d(e) \eta_0^2
\,.
\label {gw emission eq}
\end{eqnarray}

Solution without the dark wave emission (the case $f_d(e) = 0$) is given by
\begin{eqnarray}
&&
R(t) = \left(4 K_{GR} (t_e - t) \right)^{1/4}
\,, \hspace{0.5cm}
K_{GR} = \frac{64}{5} G^3 \mu M^2 \, f_g(e) = \frac{8}{5} r_S^3 \frac{\mu}{M} \, f_g(e)
\,.
\end{eqnarray}
Time scale in the last stage of gravitational collapse
is $r_S \sim 10^{-5} \mu{\rm sec} (2 GM/ 3 {\rm km})$.

To solve the problem of gravitational collapse at BH-NS merger, 
we need to derive the differential equation for dark scalar wave emission.

\vspace{0.5cm}
 {\bf Dark wave emission in isolated neutron star or white dwarf}

Consider an isolated neutron star or a white dwarf.
Neutron star is supported by the degeneracy pressure of neutrons against gravitational collapse,
while white dwarf by the electron degeneracy pressure.
This picture has been confirmed by many observations when
GR holds without the presence of eJBD field:
both objects are stable in isolation.
Stability of compact objects in isolation are guaranteed due to very small
$\chi$ coupling of gravitational strength order to matter \cite{dilaton coupling constraint}.
Nevertheless, they may emit the dark wave in dynamical process 
of  the gravitational collapse at merger with black hole.

The fundamental calculation of dark wave emission rate by \cite{dw emission rate}
gives the gravitational mass change, while we also need dark charge emission rate.
These two rates are related.
An intuitive way to convert from the energy to the charge is
to multiply the factor $D_{NS}/m_{NS}$ for this conversion.
\footnote{
The paper \cite{dw emission rate} adopts for the matter system N-body
geodesic action in which particle masses may depend
on JBD field, hence $m_i(\chi)$.
Derivative $d m_i/d\chi$ plays the fundamental role
in evaluating dipole dark wave emission rate, 
 consistent with our conversion factor $D_{NS}/m_{NS}$.
}

It is possible to justify this conversion factor another way, by adopting 
an adiabatic change of static configuration in eJBD gravity.
In this picture one takes parameters of static solutions as slowly changing functions of time.
In particular, the parameter $l$ in the static solution above, hence
its function  $M^2 + D^2 = l^2/(2G)^2$  slowly decreases in time.
Since $ \Delta ( M^2 + D^2) \sim M \Delta M + D \Delta D $,
we take this to imply that the gravitational mass variation with weight $M$
is accompanied by $D$ charge variation with weight $D$,
hence
\begin{eqnarray}
&&
\frac{d D}{dt} = \frac{D}{M_{NS}} (\frac{d M_g}{ dt})_{DW}
= -  \frac{D}{M_{NS}} L_{DW} = -G^3 \frac{M_{BH}^2 }{M_{NS}} f_d(e) \eta_0^2
 \frac{D^3 }{R^4 }
\,,
\end{eqnarray}
for BH-NS binary.

\vspace{0.5cm}
{\bf Coupled differential equations and asymptotic behavior towards gravitational collapse}

Using dimensionless variables,
\begin{eqnarray}
&&
r = \frac{R}{2G M} = \frac{R}{r_S}
\,, \hspace{0.3cm}
x = \frac{D}{M}
\,, \hspace{0.3cm}
\tau = \frac{t}{ 2 G \mu}
\,,
\end{eqnarray}
the set of differential equations to be solved is
\begin{eqnarray}
&&
\frac{dr}{d\tau} = -  \left( \frac{8}{5} f_g(e) \frac{1}{r^3} + \frac{1}{2} f_d(e) \eta_0^2
\frac{x^2}{r^2} \right) (\frac{\mu}{M})^2
\,,
\\ &&
\frac{dx}{d\tau} = - \frac{1}{8} (\frac{M_{BH} }{M} )^3 f_d(e)\eta_0^2 \, \frac{x^3}{r^4}
\,.
\end{eqnarray}

This mathematical system is not analytically tractable.
But one can attempt to derive the end point behavior and
confirm the result by numerical integrations.
We  adopt the following ansatz for solution towards the end time 
$\tau_*$ of the gravitational collapse:
\begin{eqnarray}
&&
x \sim x_* + a \ln (\tau_* - \tau) \approx x_*
\,, \hspace{0.5cm}
r \sim b (\tau_* - \tau)^{\beta}
\,.
\end{eqnarray}
Here it is assumed that time span $\tau_*-\tau$ is limited by
an inequality, $ |a \ln (\tau_* - \tau) | \leq x_*$.
Inserting these into the coupled equations, we impose consistency
of time powers and their coefficients, to derive
\begin{eqnarray}
&&
\beta = \frac{1}{4}
\,, \hspace{0.5cm}
b = (\frac{ 32}{ 5} f_g )^{1/4} \sqrt{\frac{\mu}{M}}
\,, \hspace{0.5cm}
a = \frac{5}{4} \frac{f_d }{ f_g}\eta_0^2 \,  \frac{M_{BH}^3 }{M \mu^2 }\, x_*^3
\,,
\\ &&
r(\tau) \rightarrow (\frac{32}{5} f_g \frac{\mu^2}{M^2} )^{1/4}\, (\tau_* - \tau)^{1/4}
\,.
\label {radius approximate}
\end{eqnarray}
The minimum size of remnant compact object is derived from that observers 
can watch the region outside the event horizon,
$R \geq 2G M_g$ (the Schwarzschild radius taking a mass $M_g$), 
from which one derives an equation for the minimum $M_g$:
$M_g =  M - G M \mu/(4 G M_g) $.
This gives the minimum $M_g = M (1 + \sqrt{1- \mu/M})/2 \equiv M_{\rm min}$.

\begin{figure*}[htbp]
 \begin{center}
 \epsfxsize=0.6\textwidth
 \centerline{\epsfbox{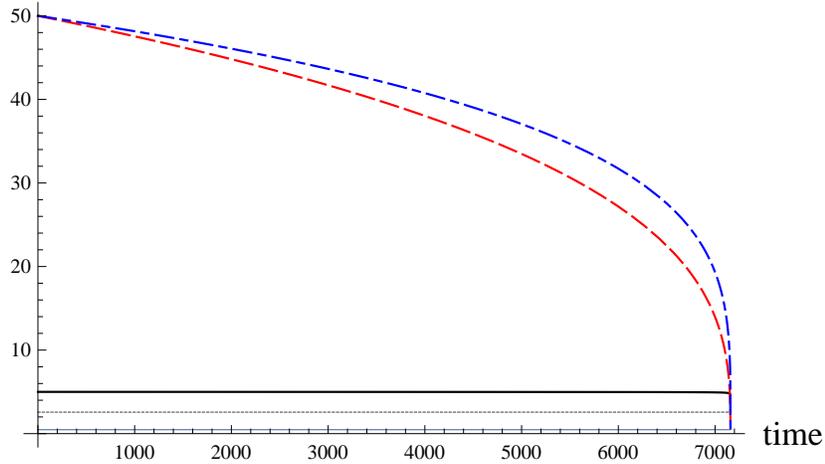}} \hspace*{\fill}\vspace*{1cm}
   \caption{
Dark charge $x$ in solid black and binary orbit radius 
$r$ in dashed orange against a rescaled time $\tau$
(to transform away $8 f_g \mu^2/5 M^2 \rightarrow 1 $ in $r-$equation)
for parameters, $\eta_0=1\,, e = 0.3\,,   m_{NS}/m_{BH} = 1.4/6$.
Assumed initial conditions are $x = 5\,, r=50$.
For comparison the GR formula of radius $\propto (\tau_*- \tau)^{1/4}$ 
is shown in dash-dotted blue, and the end point value $x$ 
and $r_*=(1+\sqrt{1-\mu/M})/4$ are in dotted black.
}
   \label {gw-dw comparison}
 \end{center} 
\end{figure*}

In Fig(\ref{gw-dw comparison}) we illustrate a
result of numerical integration, and give a comparison with
the  behavior in GR.
The mass parameters are taken from event GW200115 \cite{gw at bh-ns merger},
interpreted to be BH-NS merger event.
Computation is stopped at  the time when the orbit radius reaches the minimum value
$R = 2 G  M_{\rm min}$,
 where $x(\tau)$ is found to be positive and finite.
This indicates that haired compact object remains after gravitational collapse.

The orbit change of binary systems generates the gravitational
wave amplitude $h(\tau)$ (
metric component usually denoted
by $h^{TT}$, the transverse traceless component of metric fluctuation
from the flat Minkowski spacetime, $g_{\mu\nu} - \eta_{\mu\nu}$)
 according to \cite{gr textbook 1, gr textbook 2}
\begin{eqnarray}
&&
\frac{\dot{P}}{P} = - \frac{3}{5}  \frac{f_g(e)}{r^4}
\,, \hspace{0.3cm}
h \sim \frac{R(t_0)}{R(\tau)} \, \cos 2\pi \frac{\tau}{P}
\,,
\label {gw amp}
\end{eqnarray}
using $\tau$ variable for time.
This $h$ is taken at the emission site.
This does not include the ringdown phase caused by quasi-normal modes.
These formulas help to understand template behaviors of gravitational
wave signals in observed events \cite{gw gr-test, gw gr-test: bh-ns}.

\begin{figure*}[htbp]
 \begin{center}
 \epsfxsize=0.6\textwidth
 \centerline{\epsfbox{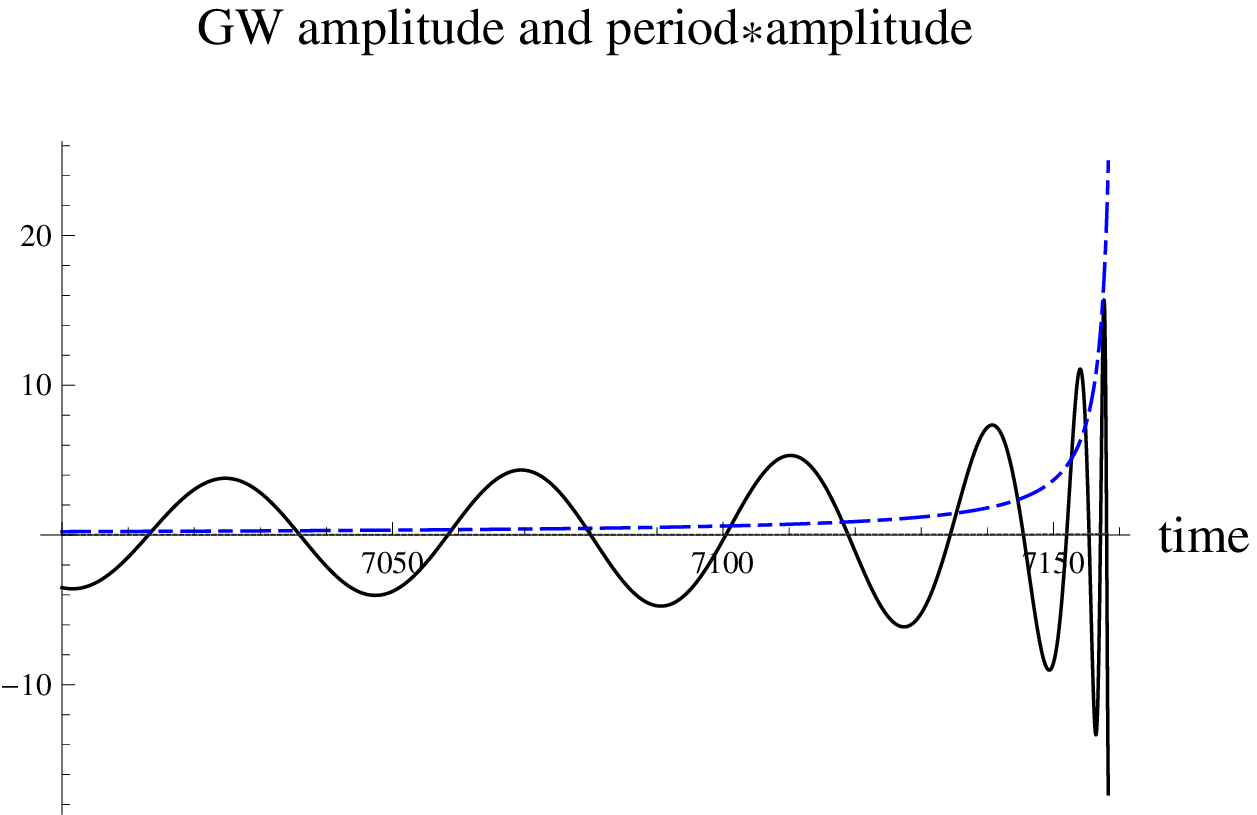}} \hspace*{\fill}\vspace*{1cm}
   \caption{
Gravitational $h-$amplitude in our unit of time $\tau$
in $\cos (2\pi \tau/P(\tau0)/r(\tau) $ for JBD gravity
in solid black and the product of period $\times $ $h-$envelope 
amplitude (scaled down by a factor $1 \times 10^{-4}/2\pi $): the product
result with dark wave emission in dotted black and result of GR in dash-dotted
blue.
The same set of parameters as in Fig(\ref{gw-dw comparison}) are used.
Computation is stopped at a time $\tau_f$
which satisfies the orbit radius $r(\tau_f)$ reaches $ 5 \times 2 GM_g(\tau_f)$
(5 times the realistic Schwarzschild radius  with calculated
$M_g(\tau_f) \sim 0.92 \times M$).
The ratio of GR to eJBD gravity
for product values changes by $\sim 290$  at the stopped time
from the initial value.
}

   \label {gw amp comparison}
 \end{center} 
\end{figure*}

Gravitational $h-$wave amplitude given by (\ref{gw amp}) is illustrated
in Fig(\ref{gw amp comparison}) using the same set of parameters as 
in Fig(\ref{gw-dw comparison}).
This example is not meant to 
give a definite conclusion on constraint of eJBD gravity,
 since a more elaborate, fully relativistic
calculation of orbit motion is required.
But it is suggestive that the parameter choice taken here seems
testable from accumulated GW observations at the BH-NS merger.
It is further found that the product of GW envelope amplitude and period 
$\propto P(\tau)/R(\tau)$
is a good measure to distinguish eJBD gravity from GR, as seen from
Fig(\ref{gw amp comparison}),
although how to match the ringdown phase is highly
non-trivial to us.

Although we present,  in our three figures, an example among many numerical computations,
the existence of remnant scalar-hair and a sharper rise of the amplitude-period product
in GR, as observed in Fig(\ref{gw amp comparison}), have been verified in
all of many other initial  combinations of the dark charge and the radius.
These two features of eJBD gravity are generic independent of initial conditions.

\vspace{0.5cm}
{\bf A remnant candidate: dark bosonic cloud}

There is no sign of naked curvature singularity at $r \geq r_*$, outside the
Shwarzschild radius, during the gravitational collapse.
In reality, the solution  must be,  for small radius $R$, matched to
solution inside a massive body of finite size.
Inside massive bodies gravity is weakened, and the body would
be shrunk until it is supported by pressure term.

There are two  possibilities for remnant compact objects
 after the gravitational collapse at the BH-NS merger.
The one is formation of black hole having a dark scalar hair,
violating the no-hair conjecture.
The other interesting  possibility for the merged remnant is the
bosonic cloud  made of black hole and eJBD scalar bosons \cite{bosonic cloud}.
\footnote{
The idea was originally proposed in \cite{axiverse}.
This a microscopic Penrose process \cite{penrose process} that extracts both energy and
angular momentum from rotating black holes.
}
This is a gigantic atom quantum mechanically bound by gravitational
interaction.
Unlike electrons around a nucleus
any number of scalar particles can occupy in a single energy level.
Hence coherent atomic transitions, superradiant graviton emission,
may be very much enhanced giving rise to characteristic GW emission rate
\cite{bosonic cloud}.
If eJBD quanta of a mass range, $10^{-20} \sim 10^{-10} $eV,
are gravitationally trapped outside the Schwarzshild radius of final BH,
its presence may become detectable.
Since eJBD field acquires a density dependent mass of order $\sqrt{G \rho_m}
\sim 1 \times 10^{-19} {\rm eV} \sqrt{\rho_m/{\rm gr\, cm}^{-3}}$
(as in \cite{chameleon}),
the quoted mass range may be realized in accretion medium surrounding a black hole. 
Thus, if the remaining dark charge $D$ is all outside the radius
$2G M_g$ of gravitational collapse,
bound eJBD quanta may all be ionized to escape from BH, 
to finally dispel dark charge consistent with the no-hair conjecture.

\vspace{0.5cm}
If experimenters find symptoms of dark wave emission from
GW detection, it may become important to find out a direct detection method of dark wave
arrival simultaneously with GW arrival on earth.
It would be further  exciting if the super-radiant GW signal is detected in the ringdown phase
after the collapse.

\subsection*{Acknowledgement}
The work is in part supported by JSPS KAKENHI Grant Nos. ~19H01899 (KK and KO),
 21H01107 (KO), and 21K03575 (MY).

\end{document}